\documentclass[12pt]{article}

\usepackage{a4wide}
\usepackage{amssymb}
\usepackage{amsmath}
\usepackage{eqnarray}
\usepackage{bbm}
\usepackage{graphicx}

\newcommand{\E}[1]{\mathop{{\rm \bf E}\!\left\{#1\right\}}\nolimits}
\newcommand{\var}[1]{\mathop{{\rm \bf var}\!\left\{#1\right\}}\nolimits}

\usepackage{array}
\begin{document}
\title{Do the Contemporary Cubature and Unscented Kalman Filtering Methods Outperform Always the Traditional Extended Kalman Filter\,?}

\author{G.Yu.~Kulikov and M.V.~Kulikova
\thanks{The authors are with CEMAT (Center for Computational and Stochastic Mathematics), Instituto Superior T\'{e}cnico, Universidade de Lisboa,
          Av. Rovisco Pais 1,  1049-001 LISBOA, Portugal; Emails:
          gkulikov@math.ist.utl.pt; maria.kulikova@ist.utl.pt}}

\date{}
\maketitle

\begin{abstract}
This brief technical note elaborates three well-known state estimators, which are used extensively in practice. These are the rather old-fashioned extended Kalman filter (EKF) and the recently-designed cubature Kalman filtering (CKF) and unscented Kalman filtering (UKF) algorithms. Nowadays, it is commonly accepted that the contemporary techniques  outperform always the traditional EKF in the accuracy of state estimation because of the higher-order approximation of the mean of propagated Gaussian density in the time- and measurement-update steps of the listed filters. However, the present paper specifies this commonly accepted opinion and shows that  despite the mentioned theoretical fact the EKF may outperform the CKF and UKF methods in the accuracy of state estimation when the stochastic system under consideration exposes a stiff behavior. That is why stiff stochastic models are difficult to deal with and require effective state estimation techniques to be designed yet.
\end{abstract}

\noindent{\bf Keywords:} Continuous-discrete stochastic state-space system, stiff model, continuous-discrete extended Kalman
filter, continuous-discrete cubature Kalman filter, continuous-discrete unscented Kalman
filter.\\

\section{Introduction}\label{sect1}

The focus of our research is the accuracy of state estimation in the so-called {\em continuous-discrete stochastic state-space} systems. This means that their process models are presented in the form of the following It$\hat{\rm o}$-type {\em Stochastic
Differential Equation} (SDE):
\begin{equation}\label{eq1.1}
dx(t)=F\bigl(t,x(t)\bigr)dt+Gdw(t),\quad t>0,
\end{equation}
where $x(t) \in \mathbb R^n$, $F:\mathbb R\times\mathbb
R^n\to\mathbb R^n $ is a nonlinear sufficiently smooth drift function, $G$ is a time-invariant
matrix of dimension $n\times q$ and $\{w(t),\,t>0\}$ is a Brownian
motion with a fixed square diffusion matrix $Q>0$ of size~$q$.  The initial state $x_0$ of SDE~(\ref{eq1.1}) is supposed to be a random variable, i.e.
$x_0 \sim {\mathcal N}(\bar x_0,\Pi_0)$ with $\Pi_0>0$, where the notation ${\mathcal N}(\bar x_0,\Pi_0)$ stands for the normal distribution with mean $\bar x_0$ and covariance $\Pi_0$. We point out that only {\em additive-noise} SDE models are studied in this paper.
At the same time, the utilized measurement models are discrete-time and given by the formula
\begin{equation}\label{eq1.2}
    z_k = h(x_{k})+v_k,   \quad k \ge 1,
\end{equation}
where $k$ stands for a discrete time index (i.e. $x_k$ means
$x(t_k)$), $h:\mathbb R^{n}\to\mathbb R^{m} $ is a sufficiently smooth function and the measurement noise is $v_k \sim {\mathcal N}(0,R_k)$ with $R_k > 0$. We remark that the {\em sampling period} (or {\em waiting time}) $\delta:=t_{k}-t_{k-1}$, i.e. when the
additional information $z_k$ comes to consideration, is assumed to
be constant, below. In addition, all realizations of the noises~$w(t)$,
$v_k$ and the initial state $x_0$ are assumed to be taken from mutually independent
Gaussian distributions. The continuous-discrete stochastic
system (\ref{eq1.1}), (\ref{eq1.2}) is a usual state estimation problem arisen in many areas
of study as diverse as target tracking, navigation,
stochastic control, chemistry and finance
\cite{AgEl05,As70,BaLi01,CrJu04,GrAn01,GrWe01,Haseltine2005,Ja70,Ok03,RoCa04,RoSa04,Sa13,Soroush1998,Si15,Wilson1998}.  Strong arguments for using the discussed mathematical model in practical estimation tasks are outlined in \cite{Sa07}.

The key issue of our research is to identify changes in performances of filters, which are core state estimation tools in practice, when they are applied for treating stiff SDE models of the form (\ref{eq1.1}). Our first state estimator is the traditional EKF designed long ago and  presented, for example, in \cite{Ja70}. Despite its simplicity and the old-fashioned nature it has been a successful filtering means in the realm of nonlinear stochastic systems for decades \cite{Goodwin,GrAn01,Ja70,Lewis,Simon2006}. Nevertheless, the first-order approximation provided by the EKF has been criticized in many studies, which have resulted in the development of more accurate UKF~\cite{Julier1995,Julier2000,VaWa01,WaVa01,Julier2002,Julier2004} and CKF \cite{Haykin2009,Haykin2010} methods.  We point out that a lot of evidence confirming the superiority of the latter filters towards the EKF in estimating various continuous- and discrete-time stochastic state-space systems have been presented in \cite{Julier1995,Julier2000,VaWa01,WaVa01,Julier2002,Julier2004,Haykin2009,Haykin2010,RoCa04,RoSa04} and in other literature. However, all published proofs in the cited papers relate to estimation of nonstiff stochastic systems and, hence, the success of the UKF and CKF for treating stiff ones is questionable. Below, we address this issue on two stochastic models whose dynamic behavior may be both nonstiff and stiff.

To estimate these nonstiff and stiff examples of the form (\ref{eq1.1}), (\ref{eq1.2}), we utilize filters designed in the frame of the so-called {\em discrete-discrete} approach \cite{Frog2012,KuKu14IEEE_TAC,KuKu14CAM}, also known as {\em linearized discretization} in \cite{GuIs96,SaSo12}. Nevertheless, all methods used below are abbreviated to: {\em Continuous-Discrete Extended Kalman Filter} (CD-EKF), {\em Continuous-Discrete Unscented Kalman Filter} (CD-UKF) and {\em Continuous-Discrete Cubature Kalman Filter} (CD-CKF), because of the continuous-discrete fashion of the state estimation task under consideration. It is also worthwhile to emphasize that the time-invariant form of the matrices $G$ and $Q$ in SDE (\ref{eq1.1}) is crucial in our research. This requirement is imposed by the CD-CKF and CD-UKF method developments in \cite{Haykin2010,KuKu16IEEE_TSP}, which are grounded in the It$\hat{\rm o}$-Taylor expansion of order~1.5 (IT-1.5). In the case of the variable matrices, the underlying stochastic discretization method IT-1.5 has a much more complicated form in comparison to that utilized in the latter papers (see further details in \cite[Sec.~10.4]{KlPl99}). Thus, the cited CD-CKF and CD-UKF methods are not applicable to such models.

\section{The State-of-the-art State Estimation Methods}\label{sect2}

In this section, we present all technical particulars of the used state estimators. They allow for an independent inspection of all calculations presented in Sec.~\ref{sect3}. We begin with the classical EKF.

\subsection{Continuous-Discrete Extended Kalman
Filter}\label{sect2a}

As said in Sec.~\ref{sect1}, all filters considered in our study are obtained within the discrete-discrete (or linearized discretization) approach and, hence, they are fixed-stepsize. The latter means that an $m$-step equidistant mesh is introduced in the sampling interval $[t_{k-1},t_k]$ of size $\delta$ for approximating SDE (\ref{eq1.1}) with the following formula:
\begin{equation}
x_{k-1}^{l+1}=x_{k-1}^{l}+\tau F\bigl(t_{k-1}^{l},x_{k-1}^{l}\bigr)+G\tilde w_{k-1}^{l}\label{eq1.3}
\end{equation}
where the mesh's nodes $t_{k-1}^{l}:=t_{k-1}+l\tau$, $l=0,1,\ldots,m$, the mesh's step size $\tau:=\delta/m$ and the discretized noise $\tilde w_{k-1}^{l} \sim {\mathcal N}(0,\tau Q)$. We point out that the utilized discretization (\ref{eq1.3}) is derived with the use of the conventional Euler-Maruyama method, which is known to be convergent of order~0.5 \cite[Sec.~10.2]{KlPl99}. Then, the subdivision number $m$ (or the step size $\tau$) plays an important role in the above discrete-time stochastic model because it is responsible for the accuracy of approximation of the original continuous-time model (\ref{eq1.1}). In other words, raising this number of steps increases the accuracy of formula (\ref{eq1.3}) and reduces the corresponding discretization error. In turn, the latter boosts the accuracy and robustness of filtering techniques grounded in this discretization and implemented for treating SDE models (\ref{eq1.1})  \cite{KuKu14IEEE_TAC,KuKu14CAM,KuKu16IEEE_TSP}.

It follows from formula~(\ref{eq1.3}) that taking the expectation yields
\begin{equation}\label{eq1.4}
\E{x_{k-1}^{l+1}}=\E{x_{k-1}^{l}}+\tau \E{F\bigl(t_{k-1}^{l},x_{k-1}^{l}\bigr)}
\end{equation}
with $\E{x_{k-1}^{l+1}}:=\hat x_{k-1}^{l+1}$ and $\E{x_{k-1}^{l}}:=\hat x_{k-1}^{l}$.
The state vector $x_{k-1}^{l}$ is independent of the noise $\tilde w_{k-1}^{l}$. Therefore the associated covariance obeys the following recursion:
\begin{equation}
\var{x_{k-1}^{l+1}}=\var{x_{k-1}^{l}+\tau F\bigl(t_{k-1}^{l},x_{k-1}^{l}\bigr)}+\tau GQG^\top.\label{eq1.5}
\end{equation}
Further, the EKF technology implies that the moment equations (\ref{eq1.4}) and  (\ref{eq1.5}) are solved approximately on each subinterval $[t_{k-1}^{l},t_{k-1}^{l+1}]$ of the Euler-Maruyama discretization (\ref{eq1.3}) \cite{Ja70}. This is done by means of the first-order Taylor expansion of the drift function $F(t_{k-1}^{l},x_{k-1}^{l})$ around the state mean vector $\hat x_{k-1}^{l}$ computed at the time $t_{k-1}^{l}$, i.e.
\begin{equation}
F\bigl(t_{k-1}^{l},x_{k-1}^{l}\!\bigr)=F\bigl(t_{k-1}^{l},\hat x_{k-1}^{l}\!\bigr)\!+\!\partial_x F\bigl(t_{k-1}^{l},\hat x_{k-1}^{l}\!\bigr)\!\bigl(x_{k-1}^{l}\!-\!\hat x_{k-1}^{l}\bigr)+HOT\label{eq1.6}
\end{equation}
where
$\partial_x F(t_{k-1}^{l},\hat x_{k-1}^{l}):=\partial F(t_{k-1}^{l},\hat x_{k-1}^{l})/\partial x_{k-1}^{l}$ is the corresponding partial derivative (Jacobian) of the function $F(t_{k-1}^{l},x_{k-1}^{l})$ with respect to $x_{k-1}^{l}$ and evaluated at $(t_{k-1}^{l},\hat x_{k-1}^{l})$, $HOT$ stands for higher-order terms of this Taylor expansion. Now neglecting $HOT$ in~(\ref{eq1.6}) and substituting it into Eqs~(\ref{eq1.4}) and  (\ref{eq1.5}) yields the classical EKF in the form of the following $m$-step state estimation algorithm.
\vspace{0.1cm}\hrule\vspace{0.1cm}
\noindent{\bf Initialization.} Set the initial state mean and covariance as follows:
$\hat x_{0|0}:= \bar x_0$, $P_{0|0}:= \Pi_0$. \\
{\bf Loop body.} For $k:=1,2,\ldots,K$, where $K$ is the number of
sampling instants in the simulation interval of SDE model (\ref{eq1.1}), do:\\
\underline{\em Time update}: Given the filtering solution $\hat x_{k-1|k-1}$ and $P_{k-1|k-1}$ at time $t_{k-1}$, compute the predicted state mean $\hat x_{k|k-1}$ and covariance matrix $P_{k|k-1}$ at the next sampling instant $t_k$. For that, set the local initial values $\hat x_{k-1|k-1}^0:=\hat x_{k-1|k-1}$ and $P_{k-1|k-1}^0:=P_{k-1|k-1}$ and fulfil the following $m$-step time-update recursive procedure with $\tau :=(t_k-t_{k-1})/m$:\\
For $l=0,1,\ldots,m-1$ do;
\begin{itemize}
\item    $t_{k-1}^{l}:=t_{k-1}+l \tau$;
\item    Evaluate the Jacobian matrix $\partial_x F(t_{k-1}^{l},\hat x_{k-1|k-1}^{l})$;
\item    $M_{k-1}^{l}:=I_{n}+\tau \partial_x F(t_{k-1}^{l},\hat x_{k-1|k-1}^{l})$, where $I_{n}$ is the identity
 matrix of size~$n$;
\item    $\hat x_{k-1|k-1}^{l+1}:=\hat x_{k-1|k-1}^{l}+\tau F\bigl(t_{k-1}^{l},\hat x_{k-1|k-1}^{l}\bigr)$;
\item    $P_{k-1|k-1}^{l+1}:=M_{k-1}^{l}P_{k-1|k-1}^{l}(M_{k-1}^{l})^\top +\tau GQG^\top.$
\end{itemize}
\underline{\em Measurement update}: Having computed the predicted state expectation $\hat x_{k|k-1}:=\hat x_{k-1|k-1}^{m}$ and the predicted covariance matrix
$P_{k|k-1}:=P_{k-1|k-1}^{m}$, one finds then the filtering solution $\hat x_{k|k}$ and
$P_{k|k}$ based on measurement $z_k$ received at the time $t_k$, as follows:
\begin{itemize}
\item Evaluate the Jacobian $\partial_x h(\hat
x_{k|k-1})$ at the state mean $\hat x_{k|k-1}$;
\item
$
 R_{e,k}:=R_k+\partial_x h(\hat
x_{k|k-1})P_{k|k-1}\partial_x h^\top(\hat
x_{k|k-1});
$
\item
$
 K_{k}:=P_{k|k-1}\partial_x h^\top(\hat
x_{k|k-1}) R_{e,k}^{-1};
$
\item
$ \hat x_{k|k}  :=   \hat x_{k|k-1}+K_{k}\left(z_k-h(\hat
x_{k|k-1})\right); $
\item
$
 P_{k|k}  := P_{k|k-1} - K_k\, \partial_x h(\hat
x_{k|k-1})P_{k|k-1}.$
\end{itemize}
\vspace{0.1cm}\hrule\vspace{0.2cm}

The presented CD-EKF calculates the {\em linear least-square estimate} $\hat x_{k|k}$  of the system's state $x(t_k)$ based on
measurements $\{z_1,\ldots,z_{k}\}$.

\subsection{Continuous-Discrete Cubature Kalman
Filter}\label{sect2b}

The CD-CKF method is invented by Arasaratnam et al. \cite{Haykin2010} and presented in the cited paper in great detail. Again, this filter is fixed-stepsize and, hence, SDE (\ref{eq1.1}) should be discretized on an equidistant mesh at first. However, Arasaratnam et al. \cite{Haykin2010} recommend the higher-order IT-1.5 discretization for constructing their CD-CKF algorithm. IT-1.5 converges with order~1.5 \cite[Sec.~10.4]{KlPl99}. That is why it is expected to provide a more accurate approximation in comparison to the Euler-Maruyama discretization (\ref{eq1.3}) on the same mesh. It results in the following discrete-time stochastic state-space model:
\begin{equation}\label{eq2.1}
x_{k-1}^{l+1}\!=\!F_d\bigl(t_{k-1}^{l},x_{k-1}^{l}\bigr)\!+\!\widetilde G w_1\!+\!{\mathbb L}F\bigl(t_{k-1}^{l},x_{k-1}^{l}\bigr) w_2
\end{equation}
with $\widetilde G:=G Q^{1/2}$, where $Q^{1/2}$ denotes the square root of $Q$, and
\begin{equation*}
F_d\bigl(t_{k-1}^{l},x_{k-1}^{l}\bigr):=x_{k-1}^{l}+\tau F\bigl(t_{k-1}^{l},x_{k-1}^{l}\bigr)+\frac{\tau^2}{2}{\mathbb L}_0 F\bigl(t_{k-1}^{l},x_{k-1}^{l}\bigr). \end{equation*}
Here, the vector $x_{k-1}^{l}$ stands for the IT-1.5 approximation to the state $x(t_{k-1}^{l})$ of SDE (\ref{eq1.1}) at the time $t_{k-1}^{l}:=t_{k-1}+l\tau$, $l=0,1,\ldots,m$, and $F(\cdot)$ is the drift function in the given SDE model. Again, $\tau:=\delta/m$ implies the step size of our subdivision of the sampling interval $[t_{k-1},t_k]$ underlying the $m$-step equidistant discretization formula (\ref{eq2.1}). The above differential operators ${\mathbb L}_0$ and ${\mathbb L}_j$ are defined as follows:
\begin{eqnarray*}
{\mathbb L}_0 &\!\!\!:=\!\!\!& \frac{\partial}{\partial t} + \sum \limits_{i = 1}^{n} F_i \frac{\partial}{\partial x_i} + \frac{1}{2}\sum \limits_{j,p,r=1}^{n} \widetilde G_{pj} \widetilde G_{rj} \frac{\partial^2}{\partial x_p \partial x_r}, \\
{\mathbb L}_j &\!\!\!:=\!\!\!& \sum \limits_{i = 1}^{n} \widetilde G_{ij} \frac{\partial}{\partial x_i}, \quad j=1,2,\ldots,n,
\end{eqnarray*}
where $\widetilde G_{ij}$ stands for the $(i,j)$-entry in the above-defined matrix $\widetilde G$. We stress that the term ${\mathbb L}F\bigl(t_{k-1}^{l},x_{k-1}^{l}\bigr)$ in discretization (\ref{eq2.1}) refers to the square matrix with each $(i,j)$-entry determined by the operator ${\mathbb L}_j F_i\bigl(t_{k-1}^{l},x_{k-1}^{l}\bigr)$. Also, the pair of correlated $n$-dimensional random Gaussian variables $w_1$ and $w_2$
is generated from the pair of uncorrelated $n$-dimensional random Gaussian variables $\nu_1$ and $\nu_2$
as follows:
$w_1:=\sqrt{\tau} \nu_1$ and $w_2:=\tau^{3/2}(\nu_1+\nu_2/\sqrt{3})/2$.

Having obtained the discretized process stochastic model~(\ref{eq2.1}), the CD-CKF technique relies on the third-degree spherical-radial cubature rule approximations of the following Gaussian integrals:
$$
\int_{\mathbb R^n} F_d\bigl(x_{k}\bigr){\mathcal N}(x_{k};\hat
x_{k|k},P_{k|k})\, dx_{k},\quad\mbox{and}\quad
\int_{\mathbb R^n} F_d\bigl(x_{k}\bigr)F_d^T\bigl(x_{k}\bigr){\mathcal N}(x_{k};\hat
x_{k|k},P_{k|k})\, dx_{k}
$$
for calculation of the mean and covariance of the propagated Gaussian density. For that, one defines the corresponding cubature nodes
\begin{equation}\label{eq2.2}
\xi_i:= \left\{
\begin{array}{ll} \sqrt{n} e_i, & i=1,2,\ldots,n,\\
            -\sqrt{n} e_{i-n}, & i=n+1,n+2,\ldots,2n,
\end{array}
\right.
\end{equation}
where $e_i$ stands for the $i$-th coordinate vector in $\mathbb R^n$ and $n$ refers to the size of SDE (\ref{eq1.1}). Then, the square-root variant of the CD-CKF presented in  \cite[Appendix~B]{Haykin2010} is implemented as follows.
\vspace{0.1cm}\hrule\vspace{0.1cm}
\noindent{\bf Initialization.} Determine the lower triangular Cholesky factor (square root) $\Pi_0^{1/2}$ of the initial state covariance matrix, i.e. $\Pi_0=\Pi_0^{1/2}\Pi_0^{\top/2}$, and the process covariance square root $Q^{1/2}$, i.e. $Q=Q^{1/2}Q^{\top/2}$. Set the initial values as follows:
$\hat x_{0|0}:= \bar x_0$, $P_{0|0}^{1/2}:= \Pi_0^{1/2}$. \\
{\bf Loop body.} For $k:=1,2,\ldots,K$, where $K$ is the number of
sampling instants in the simulation interval of SDE model (\ref{eq1.1}), do:\\
\underline{\em Time update}: Given the filtering solution $\hat x_{k-1|k-1}$ and $P_{k-1|k-1}^{1/2}$ at time $t_{k-1}$, compute the predicted state mean $\hat x_{k|k-1}$ and covariance square root $P_{k|k-1}^{1/2}$ at the next sampling instant $t_k$. For that, set the local initial values $\hat x_{k-1|k-1}^0:=\hat x_{k-1|k-1}$ and $S_{k-1|k-1}^0:=P_{k-1|k-1}^{1/2}$ and fulfil the $m$-step recursive procedure with $\tau :=(t_k-t_{k-1})/m$:\\
For $l=0,1,\ldots,m-1$ do;
\begin{itemize}
\item Create the matrix ${\cal X}^l_{k-1|k-1}:=\bigl[
\begin{array}{ccc}
{\cal X}^l_{1,k-1|k-1} & \ldots & {\cal X}^l_{2n,k-1|k-1}
\end{array}\bigr]$ of the cubature nodes
${\cal X}^l_{i,k-1|k-1}:=S^l_{k-1|k-1}\xi_i+\hat
x^l_{k-1|k-1}$, $i=1,2,\ldots,2n$, where the vectors $\xi_i$ are defined in (\ref{eq2.2});
\item Create the matrix ${\cal Y}^{l+1}_{k-1|k-1}:=\bigl[
\begin{array}{ccc}
{\cal Y}^{l+1}_{1,k-1|k-1} & \ldots & {\cal Y}^{l+1}_{2n,k-1|k-1}
\end{array}\bigr]$ of the propagated nodes ${\cal Y}^{l+1}_{i,k-1|k-1}:=F_d\bigl(t_{k-1}^{l},{\cal X}^l_{i,k-1|k-1}\bigr)$, $i=1,2,\ldots,2n$,
where the mapping $F_d(\cdot)$ is defined in (\ref{eq2.1});
\item
$\hat
x^{l+1}_{k-1|k-1}:={\cal Y}^{l+1}_{k-1|k-1} \mathbbm{1}/(2n)$, where $\mathbbm{1}$ stands for the $2n$-dimensional unitary column-vector;
\item ${\mathbb Y}^{l+1}_{k-1|k-1}:=\bigl({\cal Y}^{l+1}_{k-1|k-1}-\mathbbm{1}^\top\otimes \hat x^{l+1}_{k-1|k-1}\bigr)/\sqrt{2n}$, where $\otimes$ denotes the Kronecker tensor product (see, for instance, \cite{La69} for the definition and properties of the mentioned product; this product is also coded as the built-in function \verb"kron" in MATLAB);
\item Compute the matrix ${\mathbb F}^l_{k-1|k-1}:=\tau {\mathbb L}F_d\bigl(t_{k-1}^{l},\hat
x^l_{k-1|k-1}\bigr)/2$;
\item
$
S^{l+1}_{k-1|k-1}\!\!:=\!\left[{\mathbb Y}^{l+1}_{k-1|k-1}|\sqrt{\tau} \bigl(\widetilde G + {\mathbb F}^l_{k-1|k-1}\bigr)|\sqrt{\frac{\tau}{3}} {\mathbb F}^l_{k-1|k-1} \right]\!\Theta_l$, where $\Theta_l$ is any orthogonal rotation that lower triangularizes the right-hand
matrix and produces the square root $S^{l+1}_{k-1|k-1}$.
\end{itemize}
\underline{\em Measurement update}: Having computed the predicted state expectation $\hat x_{k|k-1}:=\hat x_{k-1|k-1}^{m}$ and the predicted covariance square root $P_{k|k-1}^{1/2}:=S^{m}_{k-1|k-1}$, one finds then the filtering solution $\hat x_{k|k}$ and
$P_{k|k}^{1/2}$ based on measurement $z_k$ received at the time $t_k$, as follows:
\begin{itemize}
\item Create the matrix ${\cal X}_{k|k-1}:=\bigl[
\begin{array}{ccc}
{\cal X}_{1,k|k-1} & \ldots & {\cal X}_{2n,k|k-1}
\end{array}\bigr]$ of the cubature nodes
${\cal X}_{i,k|k-1}\!:=\!P_{k|k-1}^{1/2} \xi_i+\hat x_{k|k-1}$, $i=1,2,\ldots,2n$;
\item Create the matrix ${\cal Z}_{k|k-1}:=\bigl[
\begin{array}{ccc}
{\cal Z}_{1,k|k-1} & \ldots & {\cal Z}_{2n,k|k-1}
\end{array}\bigr]$ of the propagated nodes
${\cal Z}_{i,k|k-1}:=h\bigl({\cal X}_{i,k|k-1}\bigr)$, $i=1,2,\ldots,2n$, where the function $h(\cdot)$ is from the measurement equation~(\ref{eq1.2});
\item
$\hat
z_{k|k-1}:={\cal Z}_{k|k-1} \mathbbm{1}/(2n)$;
\item ${\mathbb X}_{k|k-1} :=\bigl({\cal X}_{k|k-1}-\mathbbm{1}^\top\otimes \hat x_{k|k-1}\bigr)/\sqrt{2n}$;
\item ${\mathbb Z}_{k|k-1} :=\bigl({\cal Z}_{k|k-1}-\mathbbm{1}^\top\otimes \hat z_{k|k-1}\bigr)/\sqrt{2n}$;
\item Apply the Cholesky decomposition $R_k=R_k^{1/2}R_k^{T/2}$ for finding the measurement noise covariance lower triangular factor $R_k^{1/2}$;
\item Compute the cross-covariance matrix $\bar P_{xz,k}$, the square root $R_{e,k}^{1/2}$ of the innovations covariance and the square root $P_{k|k}^{1/2}$ of the filtering covariance as follows:
\begin{equation}\label{eq2.3}
\left[
\begin{array}{cc}
{\mathbb Z}_{k|k-1} & R_k^{1/2} \\
{\mathbb X}_{k|k-1}  & 0
\end{array}
\right]
\Theta_k
=
\left[
\begin{array}{cc}
R_{e,k}^{1/2} & 0 \\
\bar P_{xz,k}  & P_{k|k}^{1/2}
\end{array}
\right]
\end{equation}
where $\Theta_k$ is any orthogonal rotation that lower triangularizes the left-hand
matrix of this formula, i.e. $R_{e,k}^{1/2}$ and $P_{k|k}^{1/2}$ are lower triangular matrices. The square root $P_{k|k}^{1/2}$ of the filtering covariance matrix at the sampling time $t_k$ appears as the result of this triangularization;
\item
${\mathbb W}_{k} := \bar P_{xz,k}R_{e,k}^{-1/2}$;
\item $\hat x_{k|k} := \hat x_{k|k-1}+{\mathbb W}_{k}(z_k-\hat z_{k|k-1})$.
\end{itemize}
\vspace{0.1cm}\hrule\vspace{0.2cm}
We remark that the Cholesky decomposition of the noise covariance $R_k$  in the measurement-update step of the CD-CKF may be optional and absent in the situation when this measurement covariance matrix is time-invariant or possesses such a trivial structure that its square root is known in advance. In all such cases, the square root $R_k^{1/2}$ is set in the beginning of the filtering procedure, i.e. in the {\bf Initialization}. Our examples presented in Sec.~\ref{sect3} satisfy this condition.

\subsection{Continuous-Discrete Unscented Kalman
Filter}\label{sect2c}

The {\em Unscented Kalman Filtering} (UKF) originates from the paper of Julier et al. \cite{Julier1995}, which constructs the method for discrete-time nonlinear stochastic systems. Later on, various issues related to the UKF have been explored by many authors, including \cite{Julier2000,VaWa01,WaVa01,Julier2002,Julier2004,MeIs15a,MeIs15b} and so on. At the heart of the unscented filtering is the {\em Unscented Transform} (UT) introduced by Julier et al. \cite{Julier1995,Julier2000,Julier2002,Julier2004}. The UT implies that the set of $2n+1$ deterministically selected sigma points ${\cal X}_{i}$ (smaller sigma sets are also possible) is taken by the rule
\begin{equation}
{\cal X}_{0}=\hat x,\; {\cal X}_{i}=\hat x+\sqrt{3} P^{1/2} e_i,\; {\cal X}_{i+n}=\hat x-\sqrt{3} P^{1/2} e_i,  \label{eq2.4}
\end{equation}
$i=1,2,\ldots,n$, where, as customary, $e_i$ stands for the $i$-th coordinate vector in $\mathbb R^n$, and $P^{1/2}$ means the lower triangular Cholesky factor (square root) of the covariance matrix of a given $n$-dimensional random variable $x \sim {\mathcal N}(\hat
x,P)$, i.e. $P=P^{1/2}P^{\top/2}$. In this paper, we utilize the classical parametrization of the mentioned UT and use the following UT coefficients:
$$
w^{(m)}_0=\frac{\lambda}{n+\lambda},\quad w^{(c)}_0=\frac{\lambda}{n+\lambda}+1-\alpha^2+\beta,\quad
w^{(m)}_i=w^{(c)}_i=\frac{\lambda}{2n+2\lambda},\quad i=1,\ldots,2n,\label{eq2.5}
$$
with the fixed parameters $\alpha=1$, $\beta=0$ and $\lambda=3-n$ (see the cited papers). However, other parameterizations have also been considered in literature and an exhaustive study of this issue is published in \cite{MeIs15b}.

The sigma vectors (\ref{eq2.4}) and weights (\ref{eq2.5}) allow the mean and covariance of given Gaussian distribution to be calculated as follows:
\begin{equation}\label{eq2.6}
\hat
x=\sum_{i=0}^{2n}w^{(m)}_i {\cal X}_{i},
\quad
P=\sum_{i=0}^{2n}w^{(c)}_i ({\cal X}_{i}-\hat x)({\cal X}_{i}-\hat x)^\top.
\end{equation}
The main property of the above-defined UT is that if one changes the Gaussian distribution with the mean $\hat
x$ and covariance $P$ by a sufficiently smooth nonlinear mapping $F(\cdot)$ then the mean and covariance of the transformed random variable $F(x)$ will be calculated approximately by the same formulas (\ref{eq2.6}) but with the sigma vectors ${\cal X}_{i}$ replaced by the transformed ones $F({\cal X}_{i})$ and with the mean $\hat x$ replaced by the mean evaluated for the transformed distribution by the first formula in (\ref{eq2.6}) \cite{Julier2000,VaWa01,WaVa01,Julier2002,Julier2004}.

For treating continuous-time nonlinear stochastic systems, the continuous-discrete variant of the UKF is developed in \cite{KuKu16IEEE_TSP}. Again, it is grounded the in the stochastic IT-1.5 discretization of SDE (\ref{eq1.1}) and uses the the additive (zero-mean) noise case UKF algorithm \cite[Table~7.3]{WaVa01} for estimating the discretized process model  (\ref{eq2.1}). Eventually, one arrives at the following CD-UKF method, which is tested on nonstiff and stiff models in Sec.~\ref{sect3}, numerically.
\vspace{0.1cm}\hrule\vspace{0.1cm}
\noindent{\bf Initialization.} Set the initial state expectation and covariance and also the UT coefficients (\ref{eq2.5}) as follows:
$\hat x_{0|0}:= \bar x_0$, $P_{0|0}:= \Pi_0$, $ W_m:=\bigl[
\begin{array}{ccc}
w^{(m)}_0 & \ldots & w^{(m)}_{2n}
\end{array}\bigr]^\top$, ${\cal W}:=\bigl(I_{2n+1}-\mathbbm{1}^\top\otimes
W_m \bigr)\times\mbox{\rm diag}\left\{
\begin{array}{ccc}
w^{(c)}_0 & \ldots & w^{(c)}_{2n}
\end{array}\right\}\bigl(I_{2n+1}-\mathbbm{1}^\top\otimes
W_m \bigr)^\top$ (here, $\mathbbm{1}$ stands for the $(2n+1)$-dimensional unitary column-vector, $I_{2n+1}$ is the identity matrix of size $2n+1$, $\mbox{\rm diag}\left\{
\begin{array}{ccc}
w^{(c)}_0 & \ldots & w^{(c)}_{2n}
\end{array}\right\}$ denotes the diagonal matrix with the given entries on its main diagonal and $\otimes$ is the Kronecker tensor product coded by the function \verb"kron" in MATLAB). \\
{\bf Loop body.} For $k:=1,2,\ldots,K$, where $K$ is the number of
sampling instants in the simulation interval of SDE model (\ref{eq1.1}), do:\\
\underline{\em Time update}: Given the filtering solution $\hat x_{k-1|k-1}$ and $P_{k-1|k-1}$ at time $t_{k-1}$, compute the predicted state mean $\hat x_{k|k-1}$ and covariance matrix $P_{k|k-1}$ at the next sampling instant $t_k$. For that, set the local initial values $\hat x_{k-1|k-1}^0:=\hat x_{k-1|k-1}$ and $P_{k-1|k-1}^0:=P_{k-1|k-1}$ and fulfil the $m$-step recursive procedure with $\tau :=(t_k-t_{k-1})/m$:\\
For $l=0,1,\ldots,m-1$ do;
\begin{itemize}
\item Apply the Cholesky decomposition $P^l_{k-1|k-1}=S^l_{k-1|k-1}\times(S^l_{k-1|k-1})^\top$ for finding the lower triangular factor $S^l_{k-1|k-1}$;
\item Create the matrix ${\cal X}^l_{k-1|k-1}:=\bigl[
\begin{array}{ccc}
{\cal X}^l_{0,k-1|k-1} & \ldots & {\cal X}^l_{2n,k-1|k-1}
\end{array}\bigr]$ of the sigma points
${\cal X}^l_{i,k-1|k-1}$ computed by formulas (\ref{eq2.4}) with $P^{1/2}:= S^l_{k-1|k-1}$ and $\hat x:= \hat
x^l_{k-1|k-1}$;
\item Create the matrix ${\cal Y}^{l+1}_{k-1|k-1}:=\bigl[
\begin{array}{ccc}
{\cal Y}^{l+1}_{0,k-1|k-1} & \ldots & {\cal Y}^{l+1}_{2n,k-1|k-1}
\end{array}\bigr]$ of the propagated points ${\cal Y}^{l+1}_{i,k-1|k-1}:=F_d\bigl(t_{k-1}^{l},{\cal X}^l_{i,k-1|k-1}\bigr)$, $i=0,1,\ldots,2n$,
where the mapping $F_d(\cdot)$ is defined in (\ref{eq2.1});
\item
$\hat
x^{l+1}_{k-1|k-1}:={\cal Y}^{l+1}_{k-1|k-1} W_{m}$;
\item Compute the matrix ${\mathbb F}^l_{k-1|k-1}:=\tau {\mathbb L}F_d\bigl(t_{k-1}^{l},\hat
x^l_{k-1|k-1}\bigr)/2$;
\item
$P^{l+1}_{k-1|k-1}\!\!:=\!{\cal Y}^{l+1}_{k-1|k-1}{\cal W} ({\cal Y}^{l+1}_{k-1|k-1})^\top\!\!+\tau \bigl(\!\widetilde G+ {\mathbb F}^l_{k-1|k-1}\!\bigr)\!\bigl(\!\widetilde G + {\mathbb F}^l_{k-1|k-1}\!\bigr)^\top\!\!+\tau {\mathbb F}^l_{k-1|k-1} ({\mathbb F}^l_{k-1|k-1})^\top\!\!/3$.
\end{itemize}
\underline{\em Measurement update}: Having computed the predicted state expectation $\hat x_{k|k-1}:=\hat x_{k-1|k-1}^{m}$ and the predicted covariance matrix
$P_{k|k-1}:=P_{k-1|k-1}^{m}$, one finds then the filtering solution $\hat x_{k|k}$ and
$P_{k|k}$ based on measurement $z_k$ received at the time $t_k$, as follows:
\begin{itemize}
\item Apply the Cholesky decomposition $P_{k|k-1}=P_{k|k-1}^{1/2}P_{k|k-1}^{\top/2}$ for finding the lower triangular factor $P_{k|k-1}^{1/2}$;
\item Create the matrix ${\cal X}_{k|k-1}:=\bigl[
\begin{array}{ccc}
{\cal X}_{0,k|k-1} & \ldots & {\cal X}_{2n,k|k-1}
\end{array}\bigr]$ of the sigma points
${\cal X}_{i,k|k-1}$ computed by formulas (\ref{eq2.4}) with $P^{1/2}:= P_{k|k-1}^{1/2}$ and $\hat x:= \hat
x_{k|k-1}$;
\item Create the matrix ${\cal Z}_{k|k-1}:=\bigl[
\begin{array}{ccc}
{\cal Z}_{0,k|k-1} & \ldots & {\cal Z}_{2n,k|k-1}
\end{array}\bigr]$ of the propagated points
${\cal Z}_{i,k|k-1}:=h\bigl({\cal X}_{i,k|k-1}\bigr)$, $i=0,1,\ldots,2n$, where the function $h(\cdot)$ is from the measurement equation~(\ref{eq1.2});
\item
$\hat
z_{k|k-1}:={\cal Z}_{k|k-1}  W_{m}$,\;  $P_{zz,k|k-1}:={\cal Z}_{k|k-1}{\cal W}\, {\cal Z}_{k|k-1}^\top+R_k$;
\item $P_{xz,k|k-1}={\cal X}_{k|k-1}{\cal W}\, {\cal Z}_{k|k-1}^\top$,\;
${\mathbb W}_{k} := P_{xz,k|k-1}P_{zz,k|k-1}^{-1}$;
\item $\hat x_{k|k} := \hat x_{k|k-1}+{\mathbb W}_{k}(z_k-\hat z_{k|k-1})$;
\item $P_{k|k}=P_{k|k-1}+ {\mathbb W}_{k}P_{zz,k|k-1}{\mathbb W}_{k}^T$.
\end{itemize}
\hrule\newpage
Note that the above CD-UKF is equivalent to the CD-UKF published in \cite{KuKu16IEEE_TSP}. However, we have used S\"arkk\"a's matrix notation \cite{Sa07} for presenting the state estimator under consideration in the concise and straight forward for the MATLAB implementation form in this paper.

In conclusion, we recall that the quality of any continuous-discrete nonlinear Gaussian state estimator depends mainly on errors of two sorts, namely, on the error in capturing the model's dynamic behavior (i.e. the {\em discretization error}) and on the error in approximating the moments of Gaussian distribution (i.e. the {\em moment approximation error}) \cite{KuKu16IEEE_TSP}. These errors influence the practical performance of this or that nonlinear Kalman-like filtering technique. The above observation is supported by exhaustive numerical tests presented in the cited paper and in the studies fulfilled in \cite{Julier2000,Julier2004,Haykin2009,Haykin2010,RoCa04,RoSa04} and in many others. In the present research, the Euler-Maruyama discretization scheme used in our CD-EKF in Sec.~\ref{sect2a} is of the strong convergence order~0.5. Whereas the discretizations underlying the CD-CKF and CD-UKF are the same and based on the It$\hat{\rm o}$-Taylor approximation of the strong convergence order~1.5 in Sec.~\ref{sect2b} and~\ref{sect2c}. The latter will be obviously more accurate than the first one if all the filters are run on the same equidistant mesh \cite{KlPl99}. Next, the mean and covariance approximations use the {\em linearization} of the moment equations (\ref{eq1.4}) and  (\ref{eq1.5}) in the CD-EKF. Therefore it provides the first-order accuracy for calculation of the filtering solution. In contrast, the CD-CKF and CD-UKF are grounded in the {\em third-degree spherical-radial cubature rule} and in the {\em unscented transform}, respectively. Then, these estimation methods enjoy the third-order approximations of state means and the first-order accuracies of covariances computed in the time- and measurement-update steps of the mentioned filters. For instance, it is proven theoretically in \cite{Sa13} (or in the other cited literature). Based on the above analysis, one would expect better accuracies achieved with the CD-CKF and CD-UKF in comparison to those of the CD-EKF. That would be in line with the conclusion made in \cite{Julier1995,Julier2000,VaWa01,WaVa01,Julier2002,Julier2004,Haykin2009,Haykin2010,RoCa04,RoSa04} and in many other papers. Below, we intend for testing this guess on stochastic systems which can also expose a stiff behavior.

\begin{figure}
\begin{tabular}{cc}
\includegraphics[width = 8.0cm]{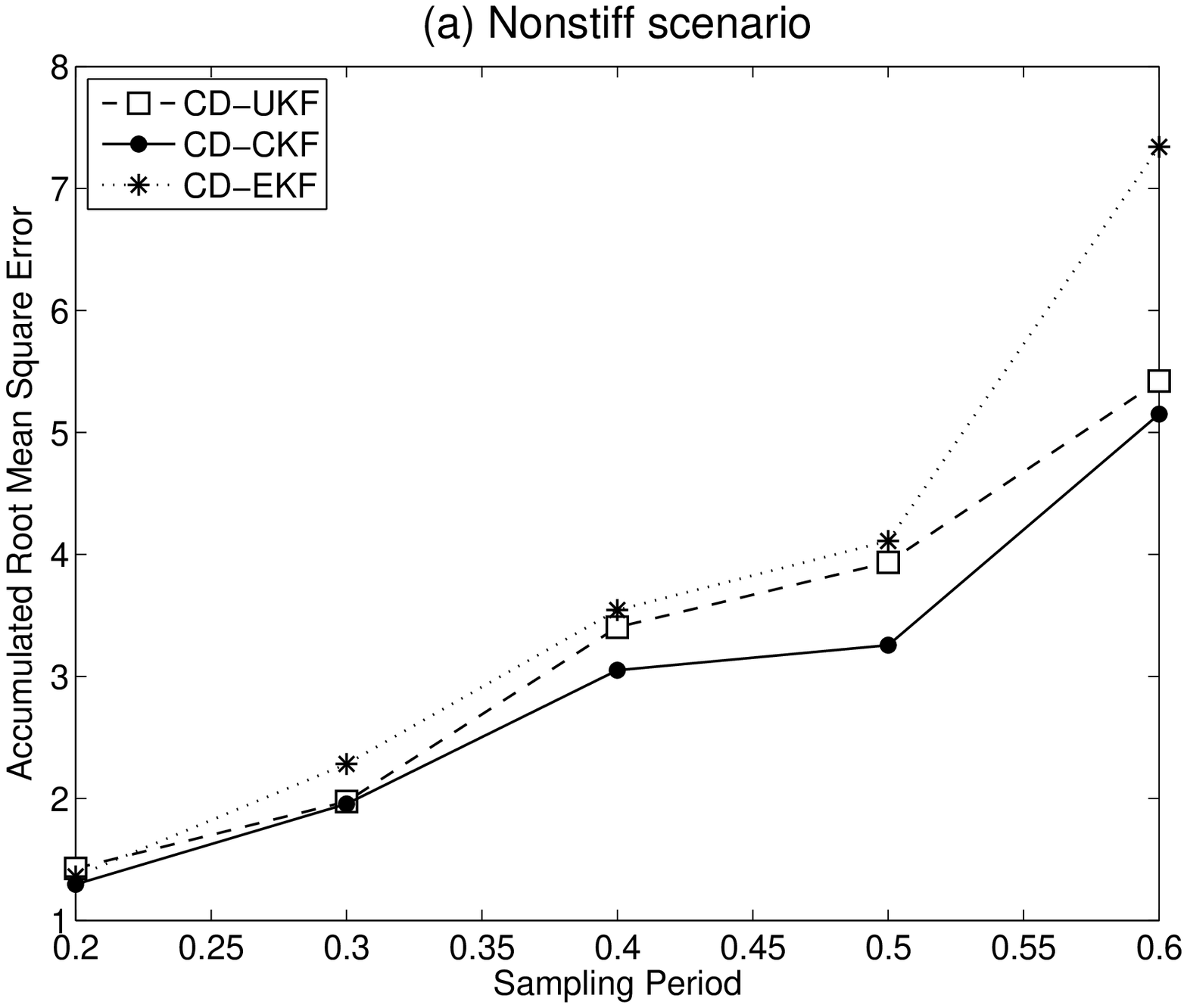} &
\includegraphics[width = 8.0cm]{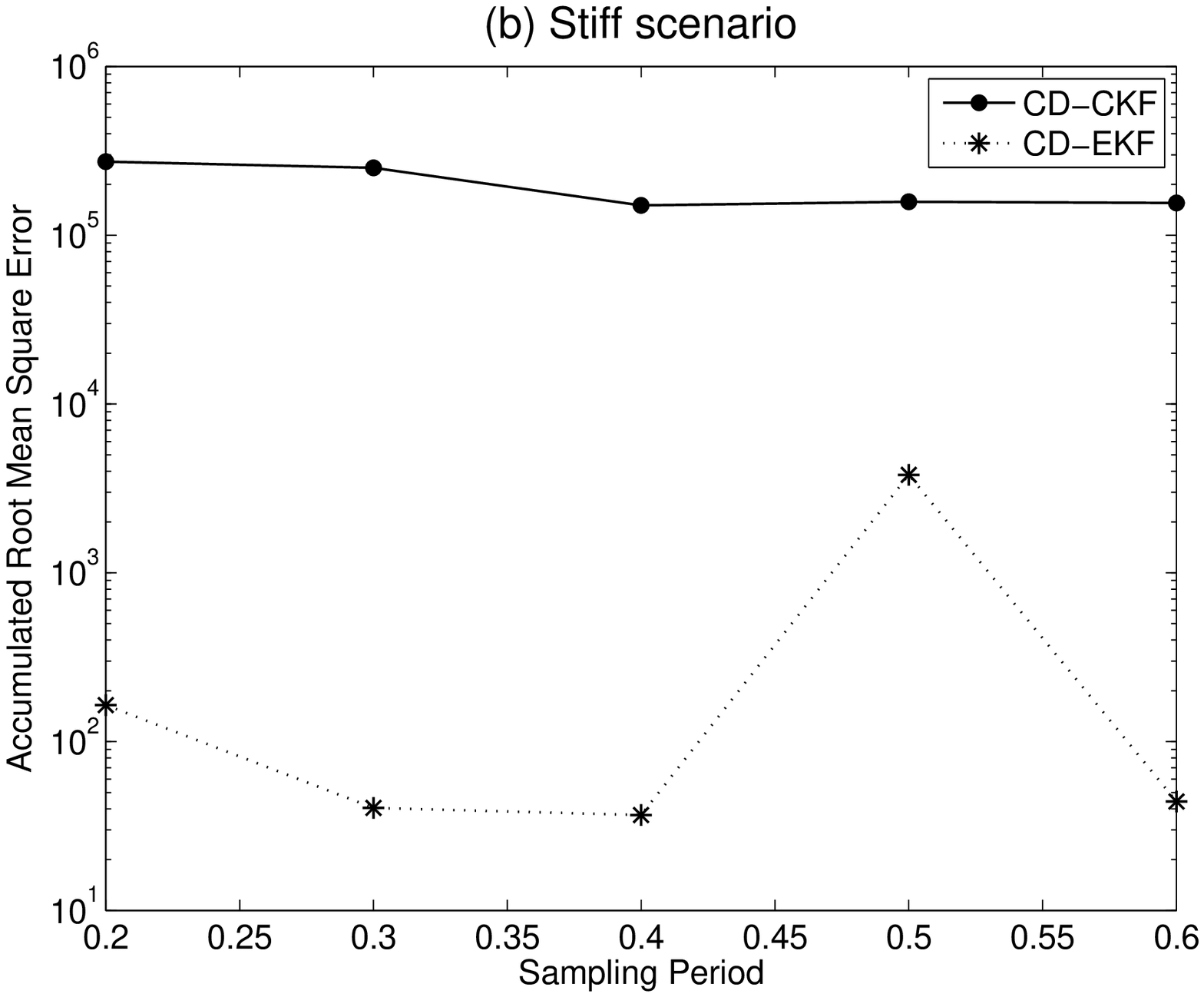}
\end{tabular}
\caption{Accuracies of the CD-EKF, CD-CKF and CD-UKF methods in the Example~3.1 scenarios} \label{fig:1}
\end{figure}

\section{Numerical Examples\label{sect3}}

Here, the CD-EKF, CD-CKF and CD-UKF methods presented in Sec.~\ref{sect2} are examined on two nonlinear stochastic systems, which can be both nonstiff and stiff depending on values of their stiffness parameters. Our intention is to study the filters' accuracies in relation to the stiffness of estimated SDE models. To assess these accuracies, we fulfil the following research. First, to simulate true states of each stochastic system used in our numerical experiments, we solve the corresponding SDE by the Euler-Maruyama  with the tiny step size equal to $10^{-5}$ in the entire simulation interval $[t_0,t_{end}]$ of the problem. Second, having fixed a sampling time $\delta$, we generate true measurements at all
$K:=[(t_{end}-t_{0})/\delta]$ sampling instants, where $[\cdot]$ stands for the
integer part of the number, by means of the computed
stochastic reference solution and the system's measurement equation. Third, we solve the reverse problem, i.e.
the given continuous-discrete stochastic system and
the simulated measurement history are treated together for yielding filtering solutions by the CD-EKF, CD-CKF and CD-UKF methods. Then, the square errors are
evaluated at the fixed sampling points. Fourth, we perform 100 Monte Carlo
runs for determining the {\em mean square error} at each
sampling instant. The accepted sampling intervals are $\delta=0.2,0.3,\ldots,0.6$ in both test examples. Thus, for all these
$\delta$'s, we compute the {\em Accumulated Root
Mean Square Error} (ARMSE)
\begin{equation}\label{eq3.1}
\mbox{\rm ARMSE}:=\Bigl[\frac{1}{100
K}\sum_{l=1}^{100}\sum_{k=1}^K\sum_{i=1}^n\bigl(xi_{ref,l}(t_k)\!-\!\widehat
{xi}_{k|k,l}\bigr)^2\!\Bigr]^{1/2}
\end{equation}
where the subscript $ref$ stands for the stochastic reference solution, $i$ indexes the estimated state vector entries in the stochastic system under consideration (i.e. $n$ is the size of the given SDE model (\ref{eq1.1})), $l$ marks the Monte Carlo
run, $k$ refers to the sampling time $t_k$ and $K$
means the total number of sampling instants $t_k$ fixed for each $\delta$. In addition, all our filters are implemented with $2\times 10^{5}$ equidistant subdivisions of each sampling interval $[t_k,t_{k-1}]$, i.e. $m=2\times 10^{5}$ in the CD-EKF, CD-CKF and CD-UKF algorithms presented in Sec.~\ref{sect2a},~\ref{sect2b} and~\ref{sect2c}, respectively. Thus, the committed discretization errors are expected to be negligible in the experiments, below.

\subsection{Van der Pol Oscillator}
\label{ex:1}

Our first test SDE model is the Van der Pol oscillator, which is considered to be a classical benchmark
by many authors (see, for instance, \cite{Frog2012,Ma08,KuKu14CAM,KuKu15RJNAMM}). However, the cited literature deals with its nonstiff variants. To cover the stiff version as well, we rescale it as explained in \cite[p.~5]{HaWa96}. This is done for estimating the Van der Pol oscillator on the same period, i.e. irrespective of its stiffness. More precisely, we consider the following process equation:
\begin{equation*}\label{eq3.2a}
d\left[
\begin{array}{c}
x1(t) \\
x2(t)
\end{array}
\right]= \left[
\begin{array}{c}
x2(t) \\
\lambda\bigl((1-x1^2(t))x2(t)-x1(t)\bigr)
\end{array}
\right]dt +\left[
\begin{array}{cc}
0 & 0 \\
0 & 1
\end{array}
\right]dw(t).
\end{equation*}
Here, $\lambda$ is the stiffness parameter and the process noise $w(t) \sim {\cal N}\left(0,I_2\right)$ where
$I_2$ stands for the identity matrix of size~2. The entire simulation
interval is taken to be $[0,2]$ in our experiment. The initial values of all the filters are fixed as follows: $\bar x_0:=[\bar x1(0),\bar x2(0)]^\top=[2,0]^\top$ and $\Pi_0={\rm diag} \{10^{-1}, 10^{-1} \}$. The formulated SDE model is observed partially, i.e. we exploit the
measurement equation
\begin{equation*}\label{eq3.2b}
z_k=x1(t_k) +v_k
\end{equation*}
with the measurement noise $v_k \sim {\cal N}\left(0,0.04\right)$.

We stress that the {\em Ordinary
Differential Equation} (ODE) underlying the Van der Pol oscillator is the known test example VDPOL for examining stiff ODE solvers \cite[p.~144]{HaWa96} and, hence, suits well to our purpose. Its stiffness depends on the value of the parameter $\lambda$. Below, we test the CD-EKF, CD-CKF and CD-UKF algorithms presented in Sec.~\ref{sect2} on the nonstiff Van der Pol oscillator with $\lambda=10$ and on the stiff one with $\lambda=10^{4}$. Our outcome is shown in Fig.~\ref{fig:1}.

Fig.~\ref{fig:1}(a) says that all the filters work well in the nonstiff Van der Pol scenario and produce more or less similar ARMSE's. However, the CD-CKF outperforms slightly the CD-UKF and more seriously the CD-EKF. This advantage is clearly seen for the longer sampling intervals $\delta$. Thus, everything is in line with the theoretical analysis presented in the concluding part of Sec.~\ref{sect2}. However, the situation changes dramatically when one estimates the stiff Van der Pol oscillator with the same methods. We stress that all the utilized data (including random sequences) and the filtering codes stay unchanged except for the value of the stiffness parameter $\lambda$ and the corresponding reference solutions. Then, Fig.~\ref{fig:1}(b) (scaled logarithmically) allows for the following conclusions. First, the accuracies of all our filters drop essentially. They become unacceptable in many cases. This is explained by the fact that the drift function $F(\cdot)$ in Example~\ref{ex:1} possesses large eigenvalues in the simulation interval $[0,2]$. In turn, such eigenvalues may enlarge the uncertainty of the state estimation and, hence, increase its ARMSE, significantly. Second, the worst method for estimating the stiff Van der Pol oscillator is the CD-UKF because it halts the performance even at the minimum $\delta =0.2$. This code reports that the calculated covariance matrix loses its positive definiteness and the MATLAB command \texttt{chol} fails to complete the demanded Cholesky factorization (that is why all ARMSE's of the CD-UKF are absent in Fig.~\ref{fig:1}(b)). The latter is not surprising because the similar CD-CKF code repots its huge ARMSE's for the same $\delta$'s. Obviously, no positive definiteness can be preserved in such circumstances, and only the square-root fashion of this method allows for completing the state estimation runs. At the same time, the CD-EKF exposes its much better performance with smaller ARMSE's. Thus, our results suggest that the CD-CKF and CD-UKF methods lose the CD-EKF in estimating stiff SDE models. Nevertheless, the accuracies of all the filters are found to be unsatisfactory within the stiff oscillator scenario. So we further confirm our observation in the reverse situation, i.e. when the CD-EKF, CD-CKF and CD-UKF are more accurate in the stiff scenario in comparison to nonstiff one.

\begin{figure}
\begin{tabular}{cc}
\includegraphics[width = 8.0cm]{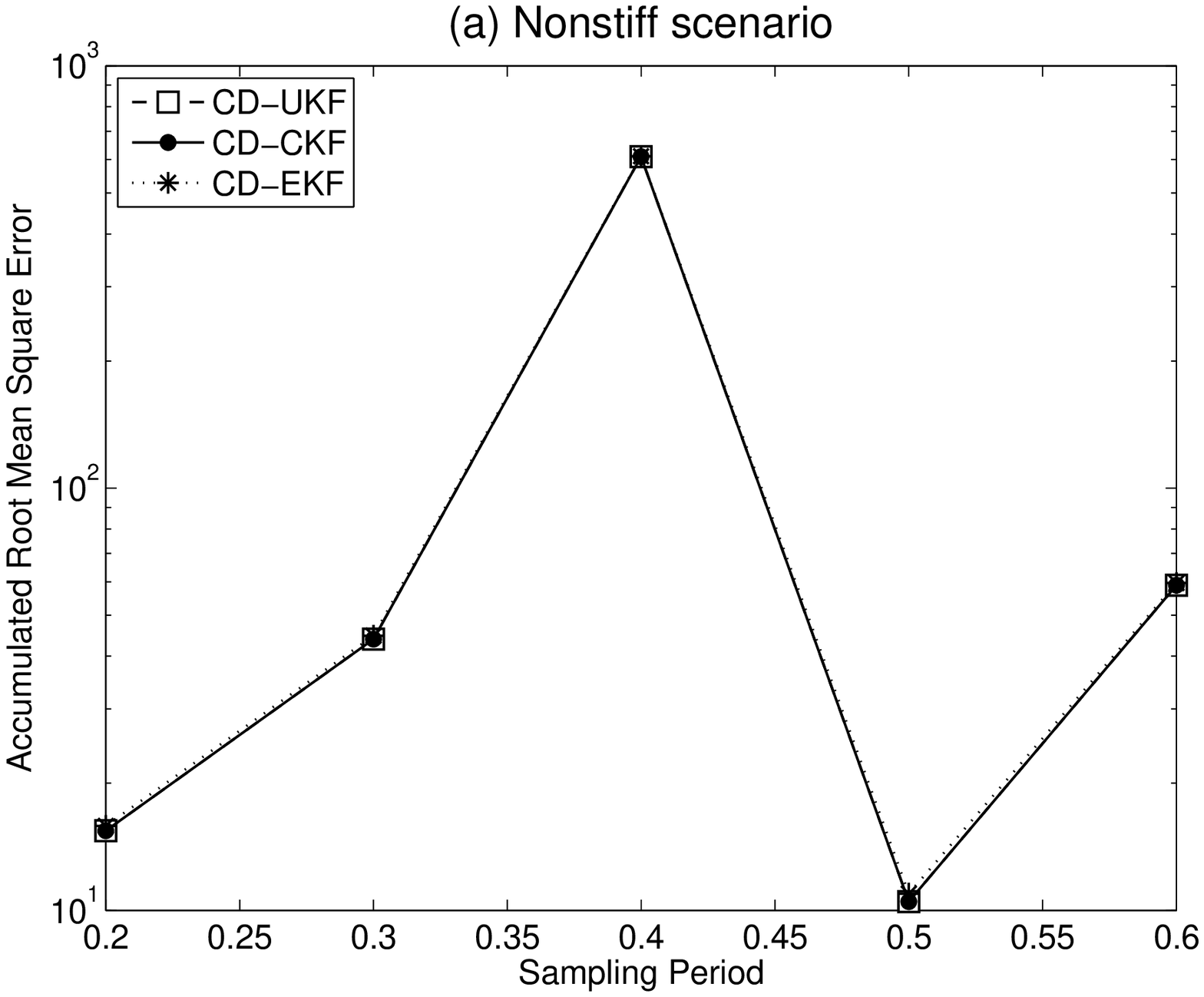} &
\includegraphics[width = 8.0cm]{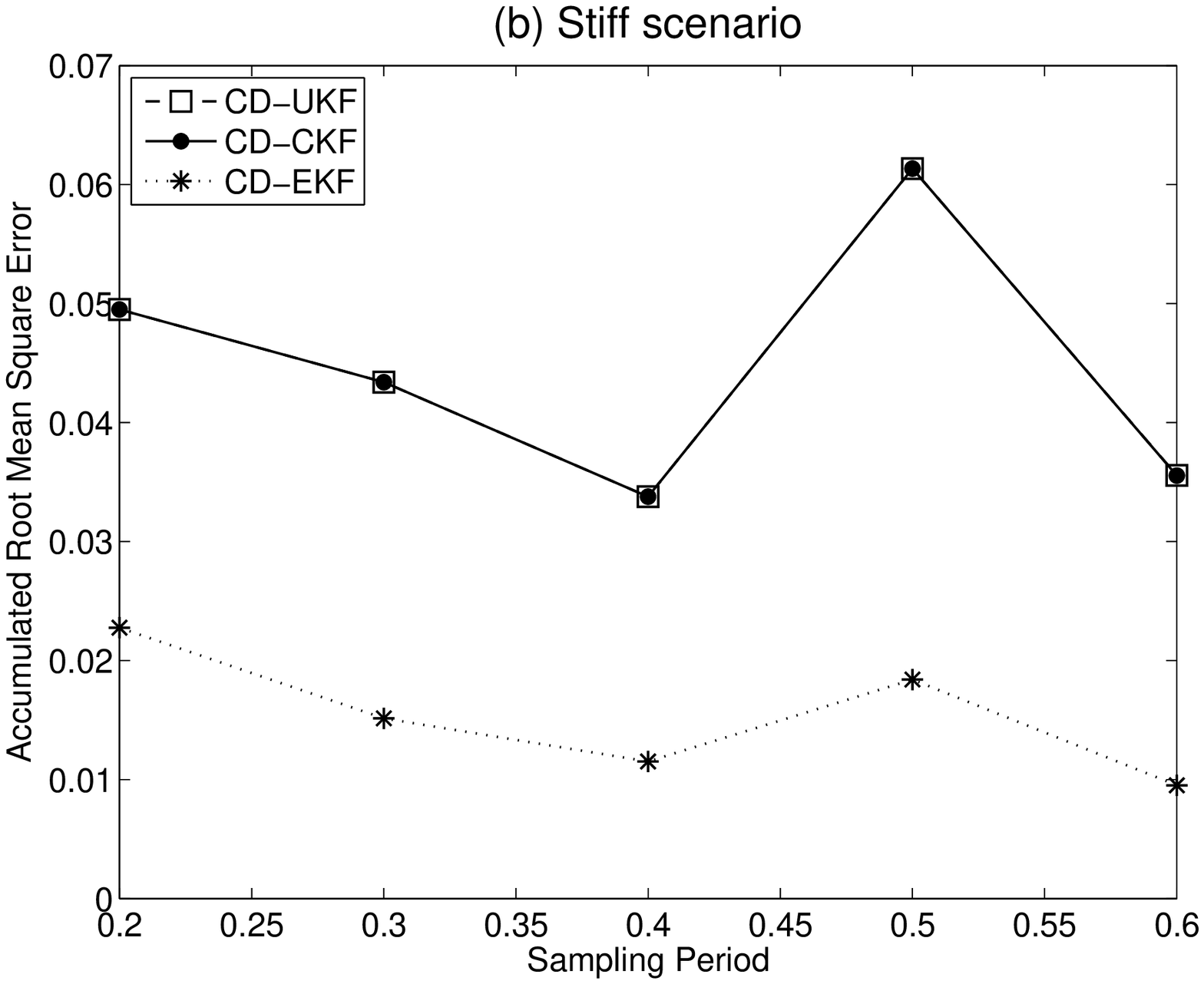}
\end{tabular}
\caption{Accuracies of the CD-EKF, CD-CKF and CD-UKF methods in the Example~3.2 scenarios} \label{fig:2}
\end{figure}

\subsection{Artificial SDE}
\label{ex:2}

Another SDE model is the continuous-discrete stochastic system already used in \cite{KuKu15EJCON,KuKu15RJNAMM}. Again, the cited papers deal with its nonstiff variant. Here, we consider the stiff version of this problem as well. We remark that the ODE underlying our SDE test is challenging and published in \cite{KuWe11CAM} in the first time. Its difficulty is also confirmed in \cite{KuWe15SISCI}, where the stiff MATLAB code \verb"ode15s" is ineffective for this ODE and is not able to find its accurate numerical solutions. The cited model is presented in the form of the following SDE:
\begin{equation*}\label{eq3.3a}
d\!\left[
\begin{array}{c}
x1(t) \\
x2(t)\\
x3(t)
\end{array}
\right]\!\!=\!\left[
\begin{array}{c} \lambda
\bigl(x2^2(t)-x1(t)\bigr)
+2{x1(t)}/{x2(t)} \\
x1(t)-x2^2(t)+1\\
-50\bigl(x2(t)-2\bigr)x3(t)
\end{array}
\right]\!dt \!+\!\left[
\begin{array}{ccc}
1 & 0 & 0 \\
0  & 0 & 0 \\
0  & 0 & 0
\end{array}
\right]\!dw(t)
\end{equation*}
where the process noise is taken to be $w \sim {\cal N}\left(0,I_3\right)$
with $I_3$ standing for the identity matrix of dimension~3. Again, this SDE
is estimated in the simulation interval $[0,2]$. The initial data of all the filters are fixed as follows: $\bar x_0:=[\bar x1(0),\bar x2(0),\bar x3(0)]^\top=[1, 1,\exp(-25)]^\top$ and $\Pi_0={\rm diag} \{10^{-4}, 10^{-4}, 10^{-4} \}$. The formulated SDE model is observed partially, i.e. we exploit the
measurement equation
\begin{equation*}\label{eq3.3b}
z_k=x2(t_k) +v_k
\end{equation*}
with the measurement noise $v_k \sim {\cal
N}\left(0,0.04\right)$.

Fig.~\ref{fig:2} exposes the outcome of our numerical simulation of Example~\ref{ex:2} in nonstiff and stiff scenarios. Our nonstiff scenario corresponds to $\lambda=10$, whereas the stiff one is obtained by increasing its stiffness to $\lambda=10^{4}$, as in Example~\ref{ex:1}. Fig.~\ref{fig:2}(a) is scaled logarithmically.

Despite its reverse matter in the accuracies of the state estimation, Example~\ref{ex:2} undoubtedly confirms our observation on the better performance of the traditional EKF in comparison to the contemporary CKF and UKF methods for estimating stiff continuous-discrete stochastic systems. This is clearly seen in Fig.~\ref{fig:2}(b). In contrast, the accuracies of all the filters under examination are in line with the commonly accepted opinion on their performances within our nonstiff scenario presented in Fig.~\ref{fig:2}(a). Here, the ARMSE's of the CD-EKF are about 2\% larger in average than those of the CD-CKF and CD-UKF. Thus, stiff continuous-discrete stochastic systems constitute a special class of state estimation problems for which the contemporary CKF and UKF are ineffective. A theoretical exploration of this unsatisfactory performance of the modern filtering techniques on stiff continuous-time stochastic models will be an interesting topic of future research.

\section{Concluding Remarks\label{conclusion}}

This paper has revealed quite an interesting and counterintuitive phenomenon of better performance of the traditional EKF in comparison to the contemporary CKF and UKF methods when applied to stochastic systems modeled by SDEs whose drift functions expose stiff behaviors. In other words, we have shown numerically that the lower-order filter outperform the higher-order methods in the accuracy of estimation of stiff continuous-discrete stochastic systems. So a theoretical justification of this phenomenon will be an interesting issue of future research in filtering theory. In addition, our numerical examples suggest that stiff stochastic systems constitute a family of SDE models which are much more difficult for accurate state estimation than traditional nonstiff ones and demand special filtering methods to be invented yet for their effective numerical treatment.

\section*{Acknowledgements}
The authors acknowledge the support from Portuguese National Funds through
the \emph{Funda\c{c}\~ao para a Ci\^encia e a Tecnologia} (FCT)\@
within project UID/Multi/04621/2013 and the \emph{Investigador~FCT~2013} programme.

\end{document}